# Observation of the quantum-anomalous-Hall insulator to Anderson insulator quantum phase transition and its scaling behavior

Cui-Zu Chang[1,2*], Weiwei Zhao[2*], Jian Li[3*], J. K. Jain[2], Chaoxing Liu[2], Jagadeesh. S. Moodera[1,4], Moses H. W. Chan[2]

[1]Francis Bitter Magnet Lab, Massachusetts Institute of Technology, Cambridge, MA 02139, USA

[2]The Center for Nanoscale Science and Department of Physics, The Pennsylvania State University, University Park, PA 16802-6300, USA

[3]Department of Physics, Princeton University, Princeton, NJ 08544, USA.

[4]Department of Physics, Massachusetts Institute of Technology, Cambridge, MA 02139, USA

Corresponding authors:  cxl56@psu.edu (C. L.) and mhc2@psu.edu (M. H. W. C.)

**Fundamental insight into the nature of the quantum phase transition from a superconductor to an insulator in two dimensions, or from one plateau to the next or to an insulator in quantum Hall effect, has been revealed through the study of its scaling behavior. Here, we report on the experimental observation of a quantum phase transition from a quantum-anomalous-Hall (QAH) insulator to an Anderson insulator in a magnetic topological insulator by tuning the chemical potential. Our experiment demonstrates the existence of scaling behavior from which we extract the critical exponent for this quantum phase transition. We expect that our work will motivate much further investigation of many properties of quantum phase transition in this new context.**

Quantum anomalous Hall (QAH) state is a topological quantum state displaying quantized Hall resistance and zero longitudinal resistance, similar to the quantum Hall (QH) state. However, while the QH effect requires a large external magnetic field, the origin of the QAH effect is the exchange interaction between electron spin and magnetic moments in ferromagnetic materials, and thus can occur even in the absence of an external magnetic field [1-3]. Ever since the discovery of the QAH effect in magnetically doped $(Bi, Sb)_2Te_3$ films[2-7], intensive research effort has focused on the global phase diagram and the novel transport properties of this effect, including non-local transport [6,8], zero Hall conductance plateau[9-11], delocalization behavior [7] and giant anisotropic magnetoresistance [12]. These experimental observations reveal an equivalence between QAH effect and QH effect in terms of their topological nature.

In the QH effect, the appearance of stable quantized Hall plateaus is intimately tied to the Anderson localization of bulk carriers of a two dimensional electron gas (2DEG) under a strong external magnetic field. The quantum phase transition (QPT) point between two Hall plateaus (known as plateau to plateau transition) is marked by the emergence of an extended state, which can be viewed, for a smoothly varying disorder, as the transition where the QH droplets of a state with one Chern number in the background of a QH state of another Chern number begin to percolate. Since this percolation process involves quantum tunneling between nearby chiral edge modes encircling individual droplets, it leads to a different critical exponent from that describing a classical percolation transition [13-15]. The QPT between an insulator and $v$=1 plateau belongs to the same universality class as the plateau to plateau transition, and serves as the prototype for the QPT studied in the present work. In the QAH effect, one may expect a similar percolation picture for chiral edge states around magnetic domains at the quantum critical point. In the experiment by Checkelsky *et al.* [7], evidence for quantum criticality and delocalization was shown while approaching the QAH state, but the scaling behavior was not explored due to the imprecise Hall quantization and the relatively large residual longitudinal resistance at zero magnetic field. In another experiment [11], Kou *et al.* studied the scaling behavior to extract the critical exponent by varying magnetic fields and temperatures for the QAH plateau to plateau transition around the coercive field. No data collapse was demonstrated in this experiment, and the analysis was complicated by the facts that the coercive field changes with temperature and the transition is influenced by the dynamics of the magnetization reversal process.

In this *Letter*, we report on the experimental observation and systematic analysis of the QPT between a QAH phase and an Anderson insulating phase in a V-doped $(Bi, Sb)_2Te_3$ film. The main difference from the previous work [11] is that the QPT is driven by a gate voltage that tunes the chemical potential. The essential physics can be illustrated qualitatively in a percolation picture, as sketched in Fig. 1a, which is closely analogous to the physics underlying the QH plateau-to-plateau transitions in disordered 2DEGs [16-18]. For a smoothly varying potential, three regions can be identified: the QAH phase when the chemical potential lies in the Zeeman exchange gap induced by ferromagnetism; the metallic phase when the chemical potential lies in the valence band; and an intermediate mixed region where QAH and metallic domains coexist. The presence of additional short-range disorder can drive the metallic phase (or metallic domains) into an Anderson insulator, resulting in topologically protected chiral edge modes at the domain-walls between the QAH and Anderson insulators. A topological phase transition occurs at the chemical potential where these domain-wall chiral edge modes percolate. In our experimental study of the transition by tuning the gate voltage, the external magnetic field is always kept at zero and the studied temperature range of 25mK to 500mK is well below ferromagnetic Curie temperature ($T_C$~23K). An advantage of this approach is that the magnetization of the film is essentially frozen and the QPT is not influenced by the dynamics of the magnetic domains. Consequently, we are able to perform a precise scaling analysis of our experiment data and extract the critical exponent for the QPT.

The sample used for this study is a 4 quintuple layer (QL) thick V-doped $(Bi,Sb)_2Te_3$ film on $SrTiO_3(111)$ substrate grown by a home-built molecular beam epitaxy (MBE) chamber with vacuum better than $5\times10^{-10}$Torr. As shown in Fig. 1b, a high-precision QAH state exists when the chemical potential is located in the Zeeman exchange gap induced by ferromagnetism [5]. We systematically tune the chemical potential away from the Zeeman exchange gap via a bottom gate $V_g$. Our measurements were carried out in a dilution refrigerator (Leiden Cryogenics, 10 mK, 9 T). A small excitation current of 0.1nA flowing through the film plane was used in the measurements to minimize heating, and the perpendicular magnetic field was used to initialize magnetization.

The standard four-terminal measurements were carried out. Initially, a magnetic field of 2T is applied perpendicular to the film plane to fully magnetize the sample at zero gate voltage, and the applied field is then reduced to zero for the measurements described below. Fig1b shows that at a zero magnetic field with gate bias $V_g$ at +120V the Hall resistance ($\rho_{yx}$) is quantized at $h/e^2$ and the longitudinal resistance ($\rho_{xx}$) drops to zero, indicating that the film is in a perfect QAH state. When the chemical potential is tuned away from the Zeeman exchange gap and towards the valence band, *i.e.*, with $V_g$ set at -200V, the Hall resistance $\rho_{yx}$ is no longer quantized (around $0.37h/e^2$) while the longitudinal resistance $\rho_{xx}$ becomes finite (around $5.8h/e^2$) at zero magnetic field, as shown in Fig 1c. The change of $\rho_{xx}$ from 0 to $5.8h/e^2$ signifies a transition from a state of dissipationless chiral edge transport to one with dissipative bulk transport.

The temperature dependences of the zero-field $\rho_{xx}$ and $\sigma_{xx}$ at different gate voltages $V_g$, as depicted in Fig. 2a, show a phase transition from a QAH insulator to an Anderson insulator induced by the chemical potential. A critical gate voltage ($V_g^c$) separates the $\rho_{xx}$ curves into two regions. For $V_g > V_g^c$, where the chemical potential lies in the exchange gap, we find that $\rho_{xx}$ decreases with lowering temperature, showing an apparent 'metallic' behavior, whereas $\sigma_{xx}$ also decreases, corresponding to an insulating behavior. This is a characteristic feature of the QAH state where both the longitudinal resistivity and the longitudinal conductivity vanish in the zero-temperature limit. For $V_g < V_g^c$, where the chemical potential lies in the valence band, $\rho_{xx}$ increases and $\sigma_{xx}$ decreases with decreasing temperature, exhibiting an insulating behavior. This insulator is a conventional Anderson insulator with trivial bulk state topology. We notice that in the Anderson insulating phase the longitudinal conductivity remains finite at the lowest temperature in our measurements, presumably due to residual bulk carriers, in contrast to the vanishing longitudinal conductivity in the QAH phase at the same temperature. A critical behavior is found at the gate voltage $V_g^c$ where the $\rho_{xx}$ approaches a constant value and $\sigma_{xx}$ saturates to a finite conductivity (around $0.5e^2/h$) as T→0. Since the QAH insulator and the Anderson insulator carry different Chern numbers, this is a topological QPT, similar to the insulator to plateau transition in the QH effect. The significance of disorder in our sample also implies an equivalence between the QPT studied in this work and that seen in the topological Anderson insulator[19].

Additional striking features of the above QPT are revealed by investigating resistivities $\rho_{xx}$ and $\rho_{yx}$ at zero magnetic field as a function of continuous gate voltages $V_g$ for different temperatures, as plotted in Figs. 2b and 2c. We find that all curves for $\rho_{xx}$ (Fig 2b) cross approximately at the gate voltage $V_g^c \approx -40 \pm 5V$, which identifies the QPT critical point. The systematic error $\pm 5V$ comes from the charging effect of the high dielectric $SrTiO_3$ substrate, which results in a hysteresis loop during gate voltage sweepings (inset in Fig. 2b). Further details about this hysteretic behavior are discussed in Supplementary Materials.

The scaling behavior and the critical exponent for the QPT can be analyzed quantitatively owing to the high quality of our data. To start with, we assume that close to the QPT critical point, $\rho_{xx}$ obeys a one-parameter scaling law: $\rho_{xx}(T, V_g) = \frac{h}{e^2} f(q)$, where $q = (V_g - V_g^c) T^{-\kappa}$ is the single parameter of the scaling function $f$. The validity of this hypothesis for our data is then examined, in a statistical sense, by evaluating $V_g^c$, $\kappa$ and their errors through the following data collapse method. In order to account for the hysteretic effect in gate voltage sweeping, we first shift all $\rho_{xx}$ curves along the $V_g$ axis such that they cross exactly at a single point with the critical longitudinal resistivity $\rho_{xx}^c$. From Fig. 2b, we estimate the reasonable range of $\rho_{xx}^c$ to be between 1.1 $h/e^2$ and 1.2 $h/e^2$, and hence we make evenly-spaced sampling in this range for the value of $\rho_{xx}^c$. Since a specific value of $\rho_{xx}^c$ occurs for different temperatures at different $V_g$, an average and a standard deviation for this set of $V_g$ can be obtained for each $\rho_{xx}^c$ (see Fig. 3b inset). Then the shifted data are collapsed to the best precision by recursively fitting the scaling function $f$ and the critical exponent $\kappa$ until convergence is achieved ([20] &

Supplementary Materials). A typical set of collapsed data for $\rho_{xx}(T, V_g)$ is plotted in Fig. 3a as a function of $\log|V_g - V_g^c| - \kappa \log T$ for $T = 25, 100, 300, 500$ mK. Both $V_g^c$ and $\kappa$ estimated at this stage depend on the particular choice of $\rho_{xx}^c$ (see Fig. 3b). As the final step, we take a weighted average of $V_g^c$ and $\kappa$ with respect to the sampling of $\rho_{xx}^c$, where the weights are given by the inverse-squared errors obtained in the previous step. This produces $V_g^c = -37.6 \pm 2.6$ V and $\kappa = 0.62 \pm 0.03$.

As exemplified in Fig. 3a, our fitting procedure consistently yields data collapse of a high precision in the vicinity of the QPT critical point. We do notice, however, that the range of our data that fall reasonably well on the collapsed curves, roughly from $-44$ V to 0 V, is very asymmetric with respect to the critical gate voltage $V_g^c$. A possible explanation of this asymmetry is that the bulk quantum well states are much closer in energy to the magnetization gap from below than from above (these have been probed through the non-local transport measurement [8]). These quantum well states extend in all dimensions of the sample, and hence are less prone to disorder compared to the surface states with which the QAH effect is directly associated. The residual conductivity in the Anderson insulating phase, as observed in the inset of Fig. 1c, is also consistent with the existence of these quantum well states. The presence of these quantum well states not only explains the conducting background on the Anderson insulator side of the data, but also obscures the scaling behavior of transport observables, as we demonstrate next with numerical simulations.

The standard finite-size scaling analysis is carried out based on a generic QAH model. Our numerical simulations serve two major purposes: (1) to corroborate the QPT observed in the experiment with an estimate of the critical exponent, and (2) to illustrate explicitly the influence of the coexisting quasi-extended quantum well states on scaling. Specifically, we start with a QAH lattice model that includes an additional trivial valence band in the clean limit, and study the localization length $\lambda$ in a disordered strip based on this model as a function of strip width $M$ and energy $E$ (see Supplementary Information for details). At a fixed disorder strength, the two-dimensional system described by this model exhibits a QPT from an Anderson insulator to a QAH insulator when the Fermi energy is tuned towards the nontrivial gap center from inside the bulk bands, as illustrated clearly in Fig. 4. This resembles the QPT observed in our experiment. In this context, the one-parameter scaling law dictates $\Lambda(M, E) = f_\Lambda[(E - E_c)M^{1/\nu}]$, where $\Lambda \equiv \lambda/M$ is the normalized localization length and $\nu$ is the critical exponent such that the localization length in two dimensions diverges as $|E - E_c|^{-\nu}$ at the QPT critical point. The strip width $M$ can be equated effectively with the inelastic scattering length that has a power-law dependence on temperature, such that $M \sim L_{\text{in}} \propto T^{-p/2}$ when $p$ is the temperature exponent of the inelastic scattering length [21]. Thus, we can identify $\nu = p/2\kappa$ in relating to the critical exponent extracted from our experimental data. In the example shown in Fig. 4, by analyzing the simulated data with the same statistical procedure as in our previous discussion, we obtain $\nu = 2.38 \pm 0.05$ and $E_c = 0.034 \pm 0.001$. Here, the estimated value of $\nu$ is universal (within error) for different simulations with varying parameters, and is consistent with those reported in the QH systems [15] or previous theoretical studies of QAH systems [22]. The theoretical value of $p$ depends on the range of $T$, the dominant scattering mechanism, presence or absence of magnetic fields, and sample properties (*e.g.* quantum well width) [23]. For disordered 2D metals in the diffusive regime and low magnetic fields, electron-electron interaction gives *p*=1[24].

In high magnetic fields, the measured value of $\kappa \approx 0.42$ is consistent with $p=2$ when $\nu$ is taken as the theoretical value $\nu = 2.38 \pm 0.05$; direct measurement of $p$ from the study of samples with varying widths have confirmed $p=2$ [25]. The value of the inelastic length exponent $p$ for our QAH system is not known, but our results suggest $p \approx 3.0$, which can in principle be verified following methods of Ref. [25]. It is worth recalling here that, even in QH effect, values of $\kappa$ ranging from 0.2-0.8 have been observed [25-28]. The experimenters have studied the issue systematically as a function of the nature of scattering (short range scattering from random alloy disorder vs. long range scattering from remote ionized impurities), and found that the "universal" value $\kappa \approx 0.42$ is achieved when the dominant scattering mechanism has a short range character [28]. This suggests that the critical regime may not have been reached in samples with long ranged disorder and the observed exponent is modified by the crossover behavior. A similarly detailed investigation of the dependence of the critical exponent on nature of disorder (short *vs.* long range) in QAH samples would also be desirable, but is outside the scope of the present study.

Nevertheless, the experimental observation of the phase transition combined with the corresponding scaling near the critical point through data collapse unambiguously demonstrates the existence of electrically tunable topological QPT in our system. In addition, because of the presence of the trivial valence band, we also notice the asymmetry of the scaling behaviors between $E > E_c$ and $E < E_c$ and have to restrict our simulation data to an asymmetric range of energy with respect to $E_c$ ($-0.07 \leq E \leq 0.2$) in order to obtain a reasonable fit, in conformance with the asymmetry observed in experiments.

The observed value of $\rho_{xx}$ at the critical point is temperature independent but close to $h/e^2$, which has been confirmed in four different samples. The universal value of $\rho_{xx}=h/e^2$ has been demonstrated in 2DEGs for the transition between a QH state and a Hall insulator [20]. It is tempting to expect similar result for QAH to insulator transition, but further study on more QAH samples will be needed to make a definite conclusion.

Finally, we note the presence of an anomalous non-monotonic behavior for the gate voltage dependence of the Hall resistivity $\rho_{yx}$ near $V_g$ ~-75V at an intermediate temperature of 25mK, as shown in Fig. 2c. Its physical origin is unclear to us but we speculate that it might originate either from inhomogeneous magnetic droplets in our system, or from the complex bulk electronic band structure of the valence band for this system. In any case, this anomaly does not affect the above analysis, which relies only on the Hall and longitudinal resistances close to the critical value of the gate voltage.

To summarize, we have observed an electric field induced topological QPT between the QAH insulator and Anderson insulator, and analyzed its critical scaling behavior. We find that the value of the longitudinal resistance $\rho_{xx}$ at the critical point is close to the quantum unit of the resistance $h/e^2$. Our results are in agreement with the theoretical simulation for the occurrence of a topological phase transition between the QAH insulator and Anderson insulator. At the transition point, one expects that proximity-induced superconductivity can easily drive the system into a chiral TSC phase[29]. Therefore, our current experiments appear to be perfectly poised to provide a good opportunity to identify chiral topological superconducting phase in the magnetic TI system.

**Acknowledgements:** The authors would like to thank H. Chen, A. MacDonald and N. Samarth for the helpful discussions. C.Z.C and J.S.M. acknowledge support from grants NSF (DMR-1207469), NSF (DMR-0819762) (MIT MRSEC), ONR (N00014-13-1-0301), and the STC Center for Integrated Quantum Materials under NSF grant DMR-1231319. W.Z. and M.H.W.C acknowledge support from NSF grants DMR-1420620 (Penn State MRSEC) and DMR-1103159. J.L. acknowledges support from grants ONR-N00014-14-1-0330, ONR-N00014-11-1-0635, and MURI-130-6082. J.K.J. acknowledges support from grant DOE (DE-SC0005042). C.X.L acknowledges the support from Office of Naval Research (Grant No. N00014-15-1-2675) and NSF grants DMR-1420620 (Penn State MRSEC).

**Figures and Figure captions:**

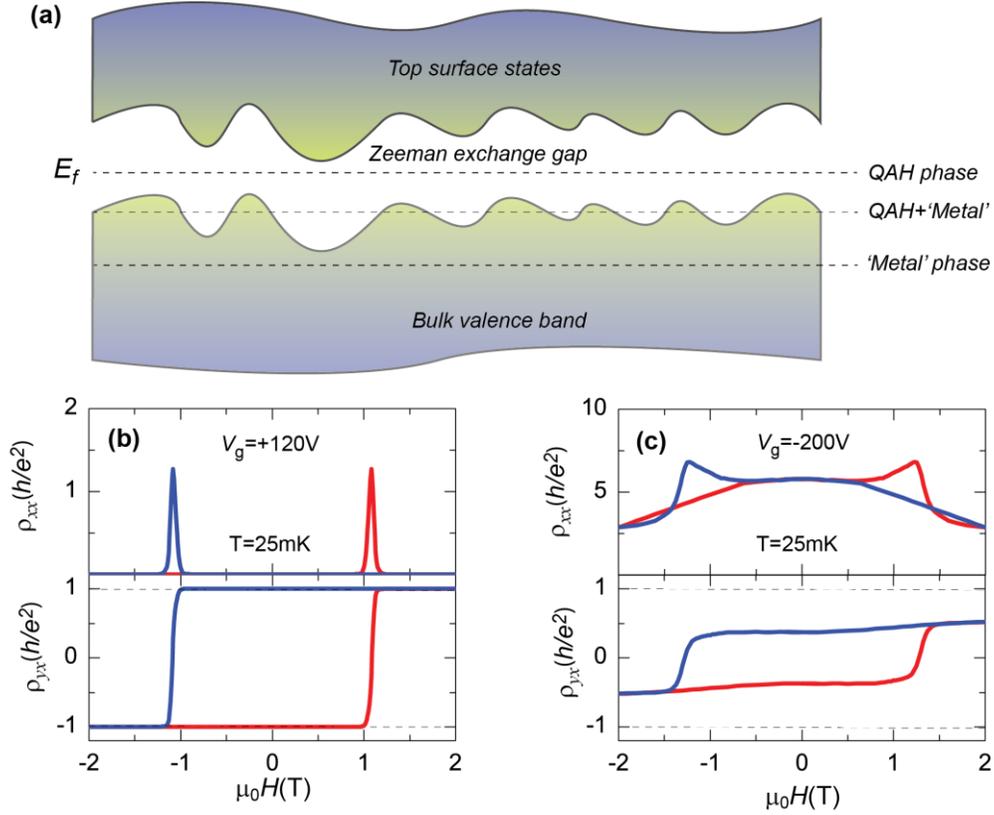

**Fig. 1** (color online) **Chemical potential dependence of topological quantum phase transition in magnetic topological insulator films.** (a) A schematic diagram depicting the QAH to Anderson insulator quantum phase transition. For a smoothly varying disorder, three regions can be identified as a function of the chemical potential: the QAH phase; the 'metal' phase; and the mixed region with coexisting QAH and metal phases. Further addition of short range disorder turns the metallic regions into an Anderson insulator. (b, c) magnetic field $\mu_0 H$ dependence of longitudinal resistance $\rho_{xx}$ and Hall resistance $\rho_{yx}$ at gate bias $V_g$=+120V (b) and $V_g$=-200V (c).

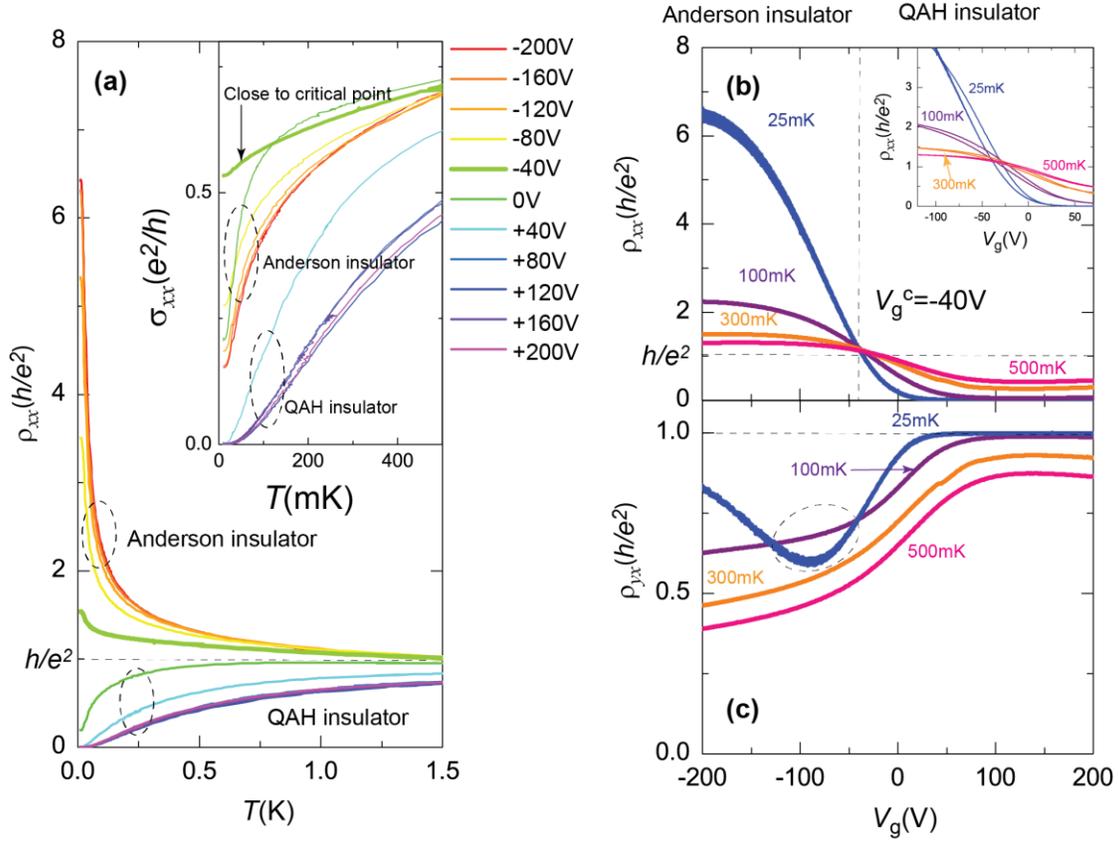

**Fig. 2** (color online) **Gate-induced topological quantum phase transition from the QAH insulator to Anderson insulator.** (a) Temperature dependence of the longitudinal resistance $\rho_{xx}$ for different $V_g$'s. Inset: temperature dependence of the longitudinal conductance $\sigma_{xx}$ for different $V_g$'s. (b, c) Gate dependence of the longitudinal resistance $\rho_{xx}$ (b) and Hall resistance $\rho_{yx}$ (c) measured at different fixed temperatures. Inset of (b) shows hysteretic behavior upon sweeping $V_g$' at different temperatures.

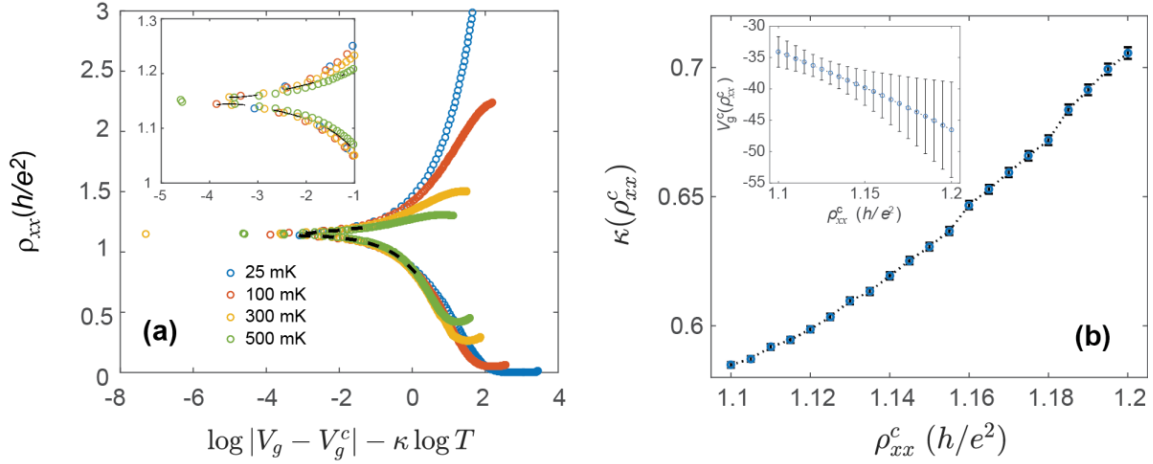

**Fig. 3** (color online) **Scaling analysis for the QAH insulator to Anderson insulator transition point.** (a) Experimental data close to the QPT critical point collapse onto the same curve with good accuracy, in agreement with the one-parameter scaling law. The inset shows a zoom-in plot of the regime close to the QPT. (b) The critical exponent $\kappa$ and the critical gate voltage $V_g^c$ (inset), as well as their errors, have been obtained by using a well-developed statistical procedure (see Supplementary Materials). The length of fitting curves (dashed lines) here also indicates the range of the experimental data used in the analysis.

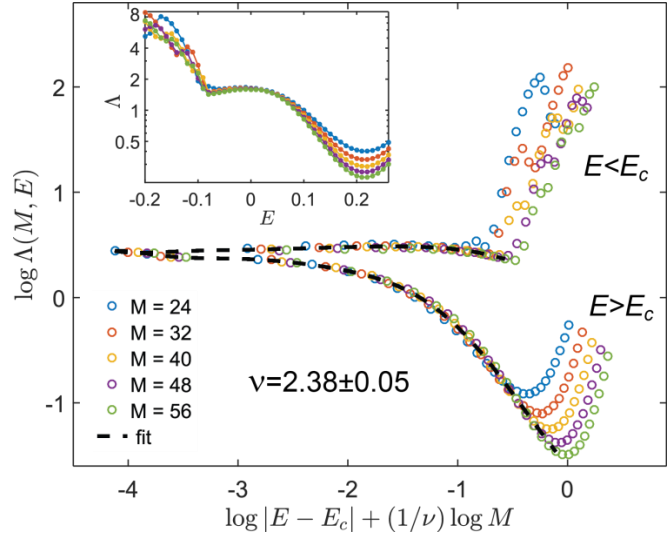

**Fig. 4** (color online) **Example of the numerical finite-size scaling analyses based on a generic QAH model** (see Supplementary Materials for details). The inset shows the normalized localization length Λ as a function of energy $E$, where the critical energy $E_c$ for the transition between an Anderson insulator ($E < E_c$) and a QAH insulator ($E > E_c$) is indicated. The normalized localization length Λ close to the QPT critical point obeys the one-parameter scaling law, as indicated by the fitting curves (dashed lines) in the main figure.